\documentclass[journal]{IEEEtran}

\usepackage{hyperref}
\usepackage{balance}
\usepackage{amsmath}
\usepackage{amssymb}
\usepackage{multirow}

\usepackage{cite}

\usepackage{ifpdf}

\ifCLASSINFOpdf
   \usepackage[pdftex]{graphicx}
   \graphicspath{{./pdf/}}
   \DeclareGraphicsExtensions{{.pdf}}
\else
   \usepackage[dvips]{graphicx}
   \graphicspath{{./eps/}}
   \DeclareGraphicsExtensions{{.eps}}
\fi

\begin{document}
\hyphenation{multi-symbol}

\title{Constellation Shaping for \\ Bit-Interleaved LDPC Coded APSK }
\author{Matthew~C.~Valenti,~\IEEEmembership{Senior~Member,~IEEE,}
        Xingyu~Xiang,~\IEEEmembership{Student Member,~IEEE,}
\thanks{Paper approved by H. Leib. Manuscript received Aug. 24, 2011; revised Feb. 13, 2012; accepted Apr. 18, 2012.}
\thanks{Portions of this paper were presented at the IEEE International Conference on Communications (ICC), Kyoto, Japan, June 2011, and the IEEE Military Communications Conference (MILCOM), Baltimore, MD, 2011.}
\thanks{This work was sponsored by the National Science Foundation under Award No. CNS-0750821 and by the United States Army Research Laboratory under Contract W911NF-10-0109.}
\thanks{M.~C.~Valenti and X.~Xiang  are with West Virginia University, Morgantown, WV, U.S.A. (email: valenti@ieee.org, xiang.xxy2008@gmail.com)).}%
\thanks{Digital Object Identifier 10.1109/TCOMM.2012.XX.XXXXXX} 
}

\maketitle


\markboth{IEEE TRANSACTIONS ON COMMUNICATIONS, ACCEPTED FOR PUBLICATION}
{VALENTI \MakeLowercase{\textit{et al.}}: Constellation Shaping for Bit-Interleaved LDPC Coded APSK}

\begin{abstract}
An energy-efficient approach is presented for shaping a bit-interleaved low-density parity-check (LDPC) coded amplitude phase-shift keying (APSK) system.  A subset of the interleaved bits output by a binary LDPC encoder are passed through a nonlinear shaping encoder whose output is more likely to be a zero than a one.  The ``shaping'' bits are used to select from among a plurality of subconstellations, while the unshaped bits are used to select the symbol within the subconstellation.  Because the shaping bits are biased, symbols from lower-energy subconstellations are selected more frequently than those from higher-energy subconstellations.  An iterative decoder shares information among the LDPC decoder, APSK demapper, and shaping decoder.  Information rates are computed for a discrete set of APSK ring radii and shaping bit probabilities, and the optimal combination of these parameters is identified for the additive white Gaussian noise (AWGN) channel.  With the assistance of extrinsic-information transfer (EXIT) charts, the degree distributions of the LDPC code are optimized for use with the shaped APSK constellation.  Simulation results show that the combination of shaping, degree-distribution optimization, and iterative decoding can achieve a gain  in excess of 1 dB in AWGN at a rate of 3 bits/symbol compared with a system that does not use shaping, uses an unoptimized code from the DVB-S2 standard, and does not iterate between decoder and demodulator.  
\vspace{-0.5cm}
\end{abstract}


\section{Introduction}
\IEEEpubidadjcol
\emph{Amplitude phase-shift keying} (APSK) is a modulation consisting of several concentric rings of signals, with each ring containing signals that are separated by a constant phase offset.  APSK has recently become widely adopted, due primarily to its inclusion in the second generation of the Digital Video Broadcasting Satellite standard, DVB-S2 \cite{dvb:2009}, where it is combined with low-density parity-check (LDPC) coding.
APSK offers an attractive combination of spectral and energy efficiency, and is especially well suited for the nonlinear channels typical of satellite systems.  

\thispagestyle{empty}

For a given modulation order $M$, an APSK constellation is characterized by the number of rings, the number of signals in each ring, the relative radii of the rings, and the phase offset of the rings relative to each other.  In \cite{gaudenzi:2006} and \cite{liolis:2008}, these parameters are optimized with an information-theoretic technique involving the maximization of the {\em symmetric information rate} (SIR), which is the mutual information between channel input and output under the assumption of equally-likely input symbols.  
The SIR is a suitable objective function when the $M$ symbols are selected with uniform probability, which is the case for 
many conventional systems, such as DVB-S2.  However, the SIR is not necessarily equal to the {\em capacity} of the channel, which is the information rate optimized over all possible input distributions.  Achieving the capacity of APSK requires a nonuniform distribution of symbols.  In \cite{liolis:2008}, a nonuniform symbol distribution is considered and optimal symbol probabilities are determined by maximizing the mutual information.  However, \cite{liolis:2008} does not specify how these distributions can be achieved in a practical system.

\begin{figure*}[t]
\centering
\includegraphics[width=5in, height=2.5in]{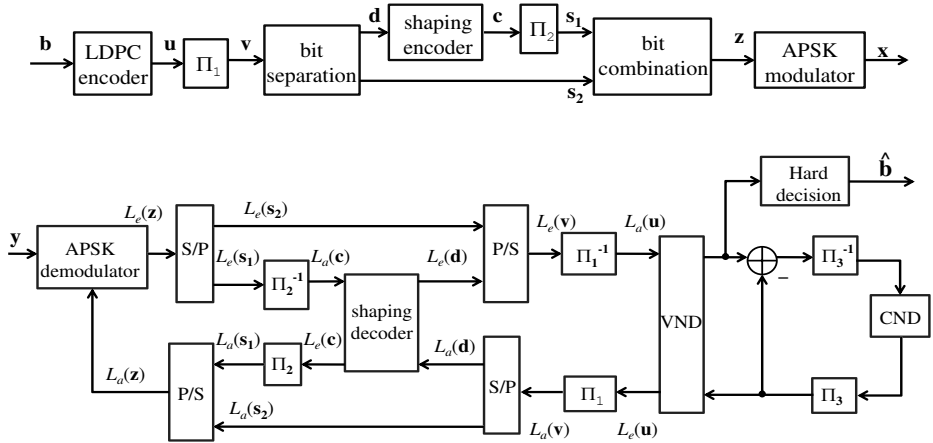}
\vspace{-0.25cm}
\caption{System model. The upper portion is the transmitter block diagram, while the lower portion is the receiver block diagram.} \label{Fig_SystemModel}
\vspace{-0.35cm}
\end{figure*}

While it is difficult to induce an arbitrary symbol distribution, {\em constellation shaping} techniques, such as the ones proposed in \cite{calderbank:1990}, can be used to bias the symbols and increase the information rate.  Constellation shaping was a topic of intense research activity in the 1980's and early 1990's.  Much of this early research activity was directed towards wireline modems, culminating in the V.34 modem standard \cite{Forney:1996}.  For a comprehensive survey of the developments leading up to the V.34 standard, the reader is directed to \cite{tretter:2002} and the references therein.  However, because wireline modems support much larger constellations than wireless systems and because this early work predates modern capacity-approaching codes, most of this early work is not readily applicable to the problem of shaping small APSK constellations for use with LDPC codes.

Recent work on constellation shaping has focused on its application to the smaller constellations typical of wireless and satellite systems.  One modern solution is to match a variable-rate turbo code with a Huffman code to allow a nonequiprobable mapping of symbols \cite{raphaeli:2004}.  A similar approach has been proposed that uses variable-rate LDPC codes \cite{Kaimalettu:2007}.  However, the requirement of a variable-rate encoder makes such solutions inconvenient.  Another solution is to use coset coding to combine turbo coding with quadrature amplitude modulation (QAM) \cite{khandani:2007}.  An approach that involves a binary-input-ternary-output turbo code and hexagonal modulation is proposed in \cite{Tanahashi:2009}.  

In \cite{legoff:2006} and \cite{legoff:2007}, Le Goff et al. propose a shaping technique suitable for bit-interleaved systems typical of modern communication standards. The technique involves the use of the short nonlinear shaping codes described in \cite{calderbank:1990} to select from among a plurality of subconstellations.  The constellation is partitioned into subconstellations such that subconstellations with lower average energy are selected more frequently than constellations with higher energy.   However, the results in \cite{legoff:2006} are limited to convolutionally coded 16-QAM, while the results in \cite{legoff:2007} are limited to turbo-coded PAM.  Neither reference considers APSK, which, as shown in this paper, is particularly well suited for constellation shaping.

In the first part of this paper, we build upon the information-theoretic optimization of APSK presented in \cite{gaudenzi:2006},\cite{liolis:2008} and the practical shaping techniques proposed in \cite{legoff:2006}, \cite{legoff:2007}.  By computing the information rates of shaped constellations, we {\em jointly} optimize parameters used by both the APSK constellation and the shaping code.
To provide compatibility with DVB-S2, we limit the APSK constellation parameters to the finite set of parameters found in the DVB-S2 standard.  Furthermore, we consider only short shaping codes in order to limit complexity.


We then turn our attention to the implementation of an actual system that uses both shaping and LDPC codes.  Initially, we limit the choice of LDPC codes to be just the codes in the DVB-S2 standard.  We then use the concept of extrinsic-information transfer (EXIT) charts \cite{tenbrink:2001,brink:2004} to optimize the LDPC code's degree distributions with respect to the shaped APSK modulation.   Our bit-error rate (BER) results show that, at a rate of 3 bits/symbol and using 32-APSK modulation, the proposed system with shaping achieves a gain of 1.13 dB relative to a more conventional DVB-S2 system.  The gain is due to three factors: (1) use of an iterative receiver, which iterates between decoder and demodulator (i.e., a so-called {\em BICM-ID} receiver \cite{li:1997,xie:vtc2009}), (2) shaping, and (3) a redesign of the LDPC code to account for the shaped-APSK modulation.  The relative gain due to each of these three factors is 0.33 dB, 0.46 dB, and 0.34 dB, respectively.  


The remainder of this paper is organized as follows.  A model for bit-interleaved coded APSK with constellation shaping is given in Section \ref{Sec_SystemModel}.  
The section describes the operation of the nonlinear shaping encoder and the receiver.  
In Section \ref{Sec_Shaping}, constellation shaping strategies are proposed for APSK constellations with mappings that are equivalent (up to a bitwise complement) to those in the DVB-S2 standard.  In Section \ref{Sec_Optimization}, the shaping and modulation parameters are selected from a finite set in order to maximize the information rate.
The implementation of an actual LDPC-coded system is presented in Section \ref{Sec_EXIT}, where EXIT-based techniques are used to optimize the LDPC degree distributions. Finally, conclusions are given in Section \ref{Sec_Conclusion}.

\section{System Model} \label{Sec_SystemModel}

\subsection{Transmitter}

A block diagram of the transmitter is shown in the upper part of Fig. \ref{Fig_SystemModel}. The system input is a length-$K_c$ vector $\mathbf b$ of equally-likely information bits, which is encoded by a rate $R_c = K_c/N_c$ binary LDPC encoder.  The length-$N_c$ codeword $\mathbf u$ at the output of the LDPC encoder is permuted by a interleaver $\Pi_1$ to generate the interleaved codeword $\mathbf v$.  A bit separator arbitrarily separates ${\mathbf v}$ into two disjoint groups, $\mathbf d$ of length $K_s$ and ${\mathbf s}_2$ of length $N_c - K_s$.   The vector $\mathbf d$ is segmented into $L$ short blocks of length $k_s$, where $K_s = L k_s$, and passed through a rate $R_s= k_s/n_s$ shaping encoder.  The shaping encoder produces zeros more frequently than ones.  The $L$ short length-$n_s$ blocks at the output of the shaping encoder are concatenated to produce the length-$N_s$ vector $\mathbf c$, where $N_s = L n_s$.  The vector $\mathbf c$ is then permuted by a second interleaver $\Pi_2$ to produce ${\mathbf s}_1$.

The vectors ${\mathbf s}_1$ and ${\mathbf s}_2$ are together used by the APSK modulator to select symbols from the constellation $\mathcal X = \{x_1, ..., x_{M}\}$, where each constellation symbol is a complex scalar.  The elements of ${\mathbf s}_1$ are said to be {\em shaping} bits, because they are used to select from among several subconstellations with a nonuniform probability.  This is in contrast with the {\em unshaped} bits ${\mathbf s}_2$, which select symbols from the selected subconstellation with uniform likelihood.  Let $g<m$ be the number of shaping bits per symbol, where $m=\log_2(M)$.  It follows that $\mathcal X$ is partitioned into $2^g$ subconstellations, each of size $2^{m-g}$.  The variables $g$ and $m$ are related to the lengths of ${\mathbf s}_1$ and ${\mathbf s}_2$ by  $N_s/(N_c-K_s) = g/(m-g)$.

Each symbol is selected from $\mathcal X$ according to the prescribed symbol mapping by using $g$ bits from ${\mathbf s}_1$ to select the subconstellation and $m-g$ bits from  ${\mathbf s}_2$ to select the symbol within the subconstellation.  The symbols in $\mathcal X$ are normalized to have average energy  $\mathcal E_s= \sum_{i=1}^{M} p(x_i) |x_i|^2$, where $p(x_i)$ is the probability that $x_i$ is selected. In Fig. \ref{Fig_SystemModel}, the two bitstreams are combined to form the vector $\mathbf z$, where each group of $m$ consecutive bits consists of $g$ bits from ${\mathbf s}_1$ and $m-g$ bits from  ${\mathbf s}_2$.   The APSK modulator uses this input and the symbol labeling map to produce a vector $\mathbf x$ of coded symbols of length $N = (N_c - K_s + N_s)/m$.

Let $p_0$ denote the probability that a particular bit in ${\mathbf s}_1$ (or, equivalently, $\mathbf c$) is equal to zero, and $p_1$ be the probability that it is equal to one.  The purpose of the shaping encoder is to produce an output with a particular $p_0>1/2$.   
The codebook is constructed such that it contains the $2^{k_s}$ distinct $n_s$-tuples of lowest possible Hamming weight.  Construction is a recursive process \cite{calderbank:1990}, with $\mathcal C$ initialized to contain the all-zeros codeword of length $n_s$.  Codewords of higher weight are recursively added to $\mathcal C$ until $|\mathcal C| = 2^{k_s}$.  Suppose that $\mathcal C$ contains all codewords of weight $w-1$ or lower.  During the next recursion, weight-$w$ codewords are repetitively drawn and added to $\mathcal C$ until either the number of distinct codeword in $\mathcal C$ is $2^{k_s}$ or all weight-$w$ codewords have been used.  In the former case, the code construction is complete, while in the latter case it moves on to begin adding codewords of weight $w+1$.

As an example, consider the $(n_s,k_s)=(9,7)$ code.  The number of codewords in $\mathcal S$ is $2^7=128$.  There are
\begin{eqnarray*}
  \binom{9}{0}
  +
  \binom{9}{1}
  +
  \binom{9}{2}
  +
  \binom{9}{3}
  & = &
  174
\end{eqnarray*}
binary 9-tuples of weight three or less.  It follows that $\mathcal S$ is a subset of these 9-tuples.  In particular, $\mathcal S$ will contain all 9-tuples of weight-2 or less, and will contain 38 of the weight-3 9-tuples.  It is preferable to select the weight-3 codewords in such a way that the column weights of matrix containing the codewords of $\mathcal C$ are balanced, thereby ensuring that $p_0$ is approximately the same in every bit position.

The overall rate of the system, $R=K_c/N$, is the number of information bits per modulated symbol, and is related to the rates of the LDPC and shaping codes by:
\begin{eqnarray}\label{rate}
  R
  & = &
  R_c
  \left[
     m + g(R_s-1)
  \right].
\end{eqnarray}
When shaping is used, $g>0$ and $R_s<1$, which implies that for a fixed $R$, the rate of the LDPC code $R_c$ with shaping must be higher than the rate of the LDPC code when shaping is not used.  

\subsection{Receiver}\label{Receiver}
The sequence of coded APSK symbols $\mathbf x$ is transmitted over an additive white Gaussian noise (AWGN) channel.  After matched-filtering, the channel output corresponding to the $k^{th}$ signaling interval is represented by the complex scalar
\begin{eqnarray}
   y_k
   & = &
   x_k
   + n_k
\end{eqnarray}
where $x_k$ is the $k^{th}$ element of $\mathbf x$, and the $\{n_k\}$ are independent and identically distributed (i.i.d.) zero-mean circularly-symmetric complex Gaussian variables with power $N_0$, which is the one-sided noise-spectral density. 
The $\{y_k\}$ corresponding to the received codeword are collected into the vector $\mathbf y$.

The receiver, which is shown in the lower part of Fig. \ref{Fig_SystemModel}, processes the received vector $\mathbf y$ and produces an estimate $\hat{\mathbf b}$ of the data bits.  The receiver uses the principle of {\em bit-interleaved coded modulation with iterative decoding} (BICM-ID) \cite{li:1997}, but includes an additional stage of processing required to decode the shaping code.  The main components of the receiver are an APSK demodulator/demapper, a shaping decoder, and an LDPC decoder,  each of which are implemented using the soft-input soft-output algorithm described below and connected by appropriate interleavers and deinterleavers. The BICM-ID receiver iterates among the demodulator, shaping decoder and LDPC decoder. The bit information exchanged in the receiver is represented as log-likelihood ratios (LLRs), following the convention that an LLR is the log of the probability of zero divided by probability of one.  The LDPC decoder consists of a variable-node decoder (VND), an edge interleaver $\Pi_3$, and a check-node decoder (CND) \cite{brink:2004}.  We use $L_a$ to denote the {\em a priori} input to a module (which may possibly be fed back from another module), and use $L_e$ to denote the output {\em extrinsic information}.


The received symbols $\mathbf y$ are passed into the APSK demodulator, which demaps the symbols and produces a vector $L_e({\mathbf z})$  containing the LLRs of the binary codeword $\mathbf z$.  The operation of the demodulator is described below in Section \ref{Section:Demod}.  By reversing the process used by the transmitter to combine ${\mathbf s}_1$ and ${\mathbf s}_2$, the LLR vector is separated by a serial-to-parallel (S/P) converter into two sequences, $L_e({\mathbf s}_1)$ and $L_e({\mathbf s}_2)$.   The sequence $L_e({\mathbf s}_1)$, which represents the LLRs of the shaping bits ${\mathbf s}_1$, is de-interleaved to produce $L_a(\mathbf c)$ and fed into the shaping decoder. The shaping decoder generates the {\em a posteriori} LLR $L_e( \mathbf d)$ by using both $L_a( \mathbf c)$ and the  {\em a priori} input information $L_a(\mathbf d)$, which is initially set to all zeros, but as described below,  will contain information fed back from the LDPC decoder in subsequent iterations.
The operation of the shaping decoder is described below in Section \ref{Section:ShapingDecoder}.  A parallel to serial (P/S) converter combines the output of the shaping decoder $L_e (\mathbf d) $ with $L_e( \mathbf s_{2})$ to produce a vector $L_e(\mathbf v)$, which represents the LLRs of vector $\mathbf v$.  The
vector $L_e(\mathbf v)$ is de-interleaved by $\Pi_1^{-1}$ and the resulting vector $L_a(\mathbf u)$ is introduced as the input to the LDPC decoder, which performs an iteration of standard sum-product decoding.

The output of the LDPC decoder is fed back to the shaping decoder and APSK demodulator to be used as extrinsic information during the next iteration. The output $L_e( \mathbf u)$ of the LDPC decoder is interleaved by $\Pi_1$.  The interleaved sequence $L_a(\mathbf v)$ is split by a S/P converter into two sequences, $L_a(\mathbf d)$ and $L_a(\mathbf s_2)$.  The vector $L_a(\mathbf d)$, which represents the {\em a priori} information on $\mathbf d$, is passed into the shaping decoder, where after the initial iteration it is used as the {\em a priori} input.  Another P/S converter, which is identical to the bit-combination block in the transmitter, combines the interleaved output of the shaping decoder $L_a(\mathbf s_1)$ with the LLRs of the unshaped bits $L_a(\mathbf s_2)$ to create the vector $L_a(\mathbf z)$, which is used as the {\em a priori} input to the demapper after the initial iteration (prior to the first iteration, $L_a( \mathbf z)$ is initialized according to the average bit probabilities for each bit position, per  Section \ref{Section:Demod}).

\subsubsection{The Demodulator}\label{Section:Demod}
The demodulator is implemented on a symbol-by-symbol basis.  For ease of exposition, we drop the dependence on the symbol interval in this subsection, so that symbols may be expressed without subscripts.  During a particular symbol interval, the demodulator computes the LLRs $L_e(\mathbf z)$ of the $m$ code bits associated with the symbol. The inputs to the demodulator are the received complex symbol $y$, which is produced by a matched-filter front end, as well as the set of $m$ {\em a priori} LLRs $L_a(\mathbf z)$, which is extrinsic information generated by the shaping and LDPC decoders during the previous iteration. Prior to the first iteration, $L_a(\mathbf z)$ is initialized to $L_a({\mathbf s}_1) = \log (\frac{p_0}{1 - p_0})$ for the shaped bits and $L_a({\mathbf s}_2)=0$ for the unshaped bits.

Let the function $f_{k}(x)$ return the $k^{th}$ bit that labels symbol $x$.  Using the MAP demodulator described in \cite{legoff:2007} and \cite{ValentiChengJSAC:2005}, the {\em a posteriori} probability that $f_{k}(x)=q, q \in \{0,1\},$ is
\begin{eqnarray}\label{demod1}
P \left( f_{k}(x) = q | y \right)
& = &
\sum\limits_{x' \in \mathcal X_k^q } p(y | x')
\mathop{ \prod_{n=1} }_{n \ne k}^{m}
\frac{ e^{f_{n}(x')L_a(z_n)}}{1 + e^{L_a(z_n)}}
\end{eqnarray}
\vspace{-0.4cm}

\noindent where $\mathcal X_k^q$ is the subset of $\mathcal X$ containing those signals whose $k^{th}$ bit position is labeled with $q$.
For the complex AWGN channel, the conditional probability of $y$ given $x$ is
\begin{eqnarray}
p(y|x)
& = &
\frac{1}{\pi N_0} \exp \left\{ -\frac{1}{N_0} |y-x|^2  \right\}.\label{awgn}
\end{eqnarray}

Expressing the output as an LLR, substituting (\ref{demod1}) and (\ref{awgn}), and canceling factors common to the numerator and denominator gives the output LLR
\begin{multline}\label{demod2}
L_e( z_k )
 =
\ln \frac{
P \left( f_{k}(x) = 0 | y \right)
}{
P \left( f_{k}(x) = 1 | y \right)
} \\
 =
\ln \frac{
\sum\limits_{x' \in \mathcal X_k^0 }
\exp \left\{
\displaystyle
- \frac{|y-x'|^2}{N_0} 
+
\displaystyle
\mathop{  \sum_{n=1} }_{n \ne k}^{m}
f_n(x')L_a(z_n) \right\}
}{
\sum\limits_{x' \in \mathcal X_k^1 }
\exp \left\{
\displaystyle
- \frac{|y-x'|^2}{N_0} 
+
\displaystyle
\mathop{  \sum_{n=1} }_{n \ne k}^{m}
f_n(x')L_a(z_n) \right\} }.
\end{multline}
The above computation may be efficiently computed using the max-star operator \cite{ValentiChengJSAC:2005}.

\subsubsection{The Shaping Decoder}\label{Section:ShapingDecoder}
The first LLR stream $L_e({\mathbf s}_1)$ is de-interleaved and fed into the shaping decoder as $L_a(\mathbf c)$. The shaping decoder outputs the extrinsic LLRs $L_e(\mathbf d)$ and $L_e(\mathbf c)$ based on the input from the demodulator and the extrinsic information fed back from the LDPC decoder.   The implementation of the shaping decoder is similar to that of the demodulator, but the summations are now over subsets of the shaping code rather than subsets of the signal constellation.  The log-likelihood of each codeword is found by taking the inner product of the $n_s$ bit LLRs with each candidate codeword.   Taking into account these differences, the output of the MAP decoder for the shaping code is
\vspace{-0.1cm}
\begin{align}\label{sdecoder1}
L_e(d_k)
 =
\ln \frac{
\displaystyle
\sum\limits_{ \mathbf d' \in \mathcal{D}_k^0}
\exp \left(
\sum\limits_{n=1}^{n_s}
f_n( \mathbf d' ) L_a(c_n)
+
\mathop{ \sum_{\ell=1} }_{\ell \ne k}^{k_s}
d'_\ell L_a(d_\ell)
\right)
}{
\displaystyle
\sum\limits_{ \mathbf d' \in \mathcal{D}_k^1}
\exp \left(
\sum\limits_{n=1}^{n_s}
f_n( \mathbf d' ) L_a(c_n)
+
\mathop{ \sum_{\ell=1} }_{\ell \ne k}^{k_s}
d'_\ell L_a(d_\ell)
\right)
}
\end{align}
where $\mathcal{D}_k^q$ denotes the set of messages $\mathbf d$ whose $k^{th}$ bit position is labeled with $q$, $q \in \{0,1\}$, and $f_n( \mathbf d' )$ is the $n^{th}$ bit in the codeword associated with message $\mathbf d'$.

The extrinsic information $L_e(\mathbf c)$ produced by the shaping decoder can be implemented in a similar manner:
\vspace{-0.1cm}
\begin{align}\label{sdecoder2}
L_e(c_k)
 =
\ln \frac{
\displaystyle
\sum\limits_{ \mathbf c' \in \mathcal{C}_k^0}
\exp \left(
\sum\limits_{n=1}^{k_s}
f_n( \mathbf c' ) L_a(d_n)
+
\mathop{ \sum_{\ell=1} }_{\ell \ne k}^{n_s}
c'_\ell L_a(c_\ell)
\right)
}{
\displaystyle
\sum\limits_{ \mathbf c' \in \mathcal{C}_k^1}
\exp \left(
\sum\limits_{n=1}^{k_s}
f_n( \mathbf c' ) L_a(d_n)
+
\mathop{ \sum_{\ell=1} }_{\ell \ne k}^{n_s}
c'_\ell L_a(c_\ell)
\right)
}
\end{align}
where $\mathcal{C}_k^q$ denotes the shaping codewords $\mathbf c$ whose $k^{th}$ bit position is labeled with $q$, $q \in \{0,1\}$, and $f_n( \mathbf c' )$ is the $n^{th}$ bit in the message associated with codeword $\mathbf c'$.  Note that, because there are more zeros than ones in the shaping codewords, $|\mathcal{C}_k^0| > |\mathcal{C}_k^1|$, and therefore there are more terms in the numerator of (\ref{sdecoder2}) than in the denominator.  This is in contrast with (\ref{sdecoder1}), which has the same number of terms in the numerator and denominator since the message bits are equally likely to be 0 or 1.

\section{Shaping Strategies} \label{Sec_Shaping}
\subsection{Shaping for 16-APSK}
\begin{figure}
\centering
\vspace{-0.25cm}
\includegraphics[width=3.4in]{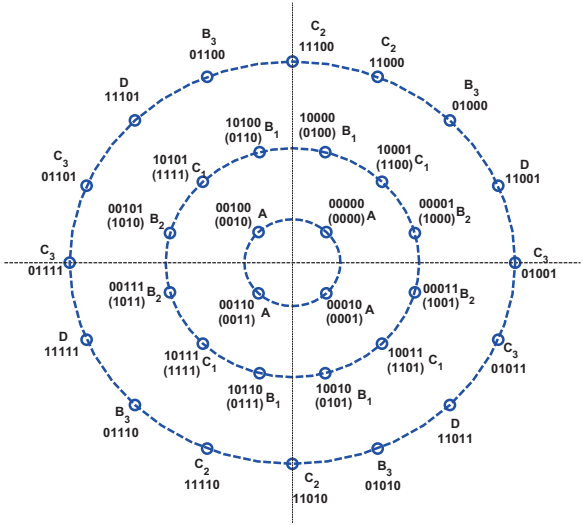}
\vspace{-0.7cm}
\caption{APSK constellations.  16-APSK uses the signals shown in the two innermost rings with symbol mappings indicated in parenthesis.   32-APSK uses all of the shown signals.} \label{Fig_APSK}
\vspace{-0.45cm}
\end{figure}

Consider the 16-APSK constellation, which uses the two innermost rings shown in Fig. \ref{Fig_APSK}.  The inner ring contains 4 symbols, while the outer ring contains 12 symbols.  The bit mapping is as indicated on the figure.  Note that this constellation is identical to the 16-APSK constellation in the DVB-S2 standard \cite{dvb:2009} and the mapping is identical except that the first two bits are complemented. The ratio of the radius of the outer ring to the radius of the inner ring is denoted $\gamma$, which according to the DVB-S2 standard may assume a value from the set $\{2.57$, $2.60$, $2.70$, $2.75$, $2.85$,  $3.15\}$.

The minimum size partition in the shaping scheme should be equal to the number of minimum-energy signals, which in this case is four.  It follows that the number of shaping bits may be either $g=1$ or $g=2$.  When $g=1$, the shaping bit is the first bit of the four-bit word labeling the constellation, and the constellation is partitioned into two subconstellations.  The first subconstellation contains symbols labeled $A$ and $B_1$, while the second subconstellation contains symbols labeled $B_2$ and $C_1$.  The first set is selected with probability $p_0$, while the second set is selected with probability $p_1$.

When $g=2$, the shaping bits are the first two bits of the word.  The signal set is partitioned into the four subconstellations indicated in Fig. \ref{Fig_APSK}; i.e., $A$, $B_1$, $B_2$, and $C_1$.  Set $A$ is selected with probability $p_0^2$, set $B_1$ and $B_2$ are each selected with probability $p_0p_1$, and set $C_1$ is selected with probability $p_1^2$.  As $p_0>p_1$, it follows that signals in set $A$ are selected most often, and signals in set $C_1$ are selected least often.

\subsection{Shaping for 32-APSK}
If all the symbols shown in Fig. \ref{Fig_APSK} are used, then the modulation is 32-APSK. The constellation consists of three concentric rings, with 4 symbols in the inner ring, 12 symbols in the middle ring, and 16 symbols in the outer ring.  This constellation is identical to the 32-APSK constellation in the DVB-S2 standard \cite{dvb:2009} and the mapping is identical except that the first and last bits are complemented.  The ratio of the radius of the middle ring to the radius of the inner ring is denoted $\gamma_1$, while the ratio of the radius of the outer ring to the radius of the inner ring is denoted $\gamma_2$.  According to the standard, the value of $\gamma = \{ \gamma_1, \gamma_2 \}$ must be one of the following: $\{ 2.53, 4.30 \}$, $\{ 2.54, 4.33 \}$, $\{ 2.64, 4.64 \}$, $\{ 2.72, 4.87 \}$, or $\{ 2.84, 5.27 \}$.

As with 16-APSK, the number of minimum energy signals is four, which is the size of the smallest partition.  The number of shaping bits should be no more than three. When $g=1$, the shaping bit is the second bit of the word.  This divides the constellation into two partitions.  The first partition, which is selected with probability $p_0$, contains the signals in the first two rings (i.e., sets $A$, $B_1$, $B_2$, and $C_1$).  The second partition, which is selected with probability $p_1$, contains the signals in the outer ring (i.e., sets $B_3$, $C_2$, $C_3$, and $D$).  With this scheme, signals in the inner two rings are more likely to be selected than signals in the outer ring.

When $g=2$, the shaping bits are the second and last bits of the word.  The signal set is partitioned into four sets: $\{A,B_1\}$, $\{B_2,C_1\}$, $\{B_3,C_2\}$, and $\{C_3,D\}$.  The partition $\{A,B_1\}$ is selected with probability $p_0^2$, making it most likely to be selected.  The partition $\{C_3,D\}$ is selected with probability $p_1^2$, making it least likely to be selected.  The other two partitions are each selected with probability $p_0p_1$.

When $g=3$, the shaping bits are the first, second, and last bits of the word. The signal set is partitioned into the eight sets indicated in Fig. \ref{Fig_APSK}; i.e., $A$, $B_1$, $B_2$, $B_3$, $C_1$, $C_2$, $C_3$, and $D$. Partition $A$ is most likely and is selected with probability $p_0^3$.  The $B_k$ partitions are selected with probability $p_0^2 p_1$, while the $C_k$ partitions are selected with probability $p_0 p_1^2$.  Finally, the $D$ partition is selected with probability $p_1^3$, which makes it least likely to be selected.

\section{Joint Parameter Optimization}\label{Sec_Optimization}
With the shaping techniques described in the previous section, energy may be conserved by using larger values of $p_0$.  However, larger values of $p_0$ generally require lower $R_s$.  To maintain a fixed overall rate $R$, using a lower $R_s$ requires a larger value of $R_c$, which weakens the effectiveness of the LDPC code.  Clearly there is a tradeoff between $R_s$ and $R_c$. To determine the optimal tradeoff, we turn to information theory.

\subsection{Optimization with Respect to CM Capacity}
In performing the optimization, we wish to determine achievable information rates by computing the mutual information between the channel input and output.  On the one hand, we could compute the mutual information between the modulated {\em symbols} at the channel input and the corresponding output.  This is called the coded modulation (CM) capacity in \cite{Caire:1998}. On the other hand, we could compute the mutual information between the {\em bits} at the input of the modulator and output of the demodulator.  In the case of a shaped system, the modulator would include the shaping encoder.  This type of capacity is called the BICM capacity in \cite{Caire:1998}.   The distinction between the two is that the BICM capacity is relevant for a system that uses a BICM receiver; i.e., doesn't feed back information from decoder to demodulator.  However, our system uses a BICM-ID receiver and is therefore able to outperform a BICM receiver and achieve performance close to that of CM.  Therefore, we will perform the optimization with respect to the CM capacity.  

Let $Y$ be the output of an AWGN channel with complex scalar input $X \in \mathcal X$.  The capacity of the channel is \cite{gallager:1968}
\begin{eqnarray}
  C
  & = &
  \max_{p \left( x \right)} I \left( X; Y \right), \label{shannoncap}
\end{eqnarray}
where $p( x)$ is the probability mass function (pmf) of $X$, the information rate (also called {\em average mutual information}) is
\begin{eqnarray}
  I \left( X; Y \right)
  & = &
  E[ i\left( X; Y \right) ], \label{expectation}
\end{eqnarray}
\vspace{-0.6cm}

\noindent and

\vspace{-0.5cm}
\begin{eqnarray}
   i \left( x; y \right)
   & = &
   \log \frac{ p \left( x, y \right) }{ p \left( x \right) p \left( y \right) } =
  \log \frac{ p \left( y | x  \right) }{ p \left( y \right) }. \label{mi}
\end{eqnarray}
Note that the expectation in (\ref{expectation}) is with respect to the joint pdf $p \left( x, y \right)$. When a base-2 logarithm is used, then (\ref{shannoncap}) has units of bits per channel use (bpcu).  

From the theorem on total probability,
\begin{eqnarray}
   p(y)
   &  = &
   \sum_{ x' \in \mathcal X} p( y|x') p ( x'). \label{py}
\end{eqnarray}
Substituting ($\ref{py}$) into ($\ref{mi}$) gives
\begin{eqnarray}
   i \left( x;  y \right)
   & = &
   \log p( y | x ) - \log \left( \sum_{x' \in \mathcal X} p(y|x') p (x') \right). \label{mi2}
\end{eqnarray}
where $p(y|x)$ is given by (\ref{awgn}) for the AWGN channel.

The capacity is found by maximizing the information rate with respect to the distribution of $X$.  In general, the optimization requires that the probability of occurrence of each of the symbols be independently varied.  Techniques for finding the optimizing distributions may be readily found in the literature, such as the Blahut-Arimoto algorithm \cite{blahut:1972}.  However, the shaping scheme proposed in Section \ref{Sec_Shaping} does not allow for independent symbol probabilities.  Rather, the symbol probabilities are related by the value of $p_0$, the number of shaping bits, and the symbol labeling.

Rather than allowing arbitrary $p(x)$, we optimize our system under the constraint of the shaping techniques described in Section  \ref{Sec_Shaping}.  For a particular constellation $\mathcal X$, number of shaping bits $g$, and $(n_s,k_s)$ shaping code, we numerically evaluate the information rate $I( X;Y )$.  While the evaluation could be done using a Monte Carlo integration \cite{Caire:1998}, we use the Gauss-Hermite quadratures method of evaluation described in \cite{philip:1995}.

\begin{figure}
\centering
\includegraphics[width=3.4in]{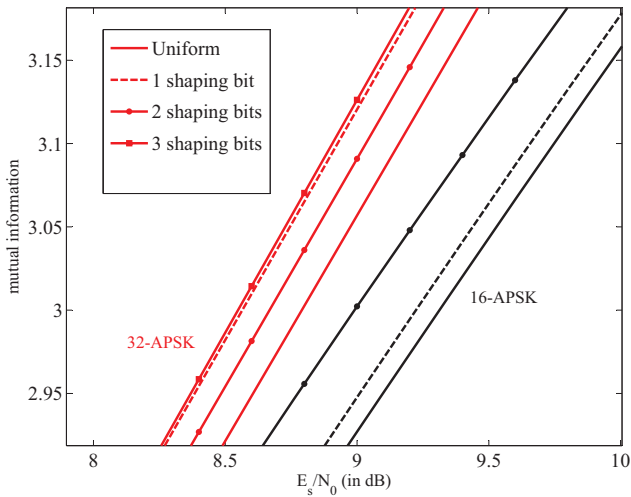}
\vspace{-0.45cm}
\caption{Information rate (in bpcu) of 16-APSK and 32-APSK over an AWGN channel.  The number of shaping bits used is indicated.  The solid line on the right of each group corresponds to a uniform input distribution (no shaping).   The rates are maximized over the permissible ring-radius ratios. } \label{Fig_Apsk16_32Cap}
\vspace{-0.5cm}
\end{figure}

We begin by evaluating the information rate under the assumption that the input symbols have a uniform distribution; i.e., the symmetric information rate.  For each $M$, we limit the ring-radii of the APSK constellation to be chosen from among the values specified in the DVB-S2 standard.   We further limit the inter-ring phase offsets to be the same as in DVB-S2, as the phase offsets have been shown to have a negligible effect on the information rate \cite{gaudenzi:2006}.  The symmetric information rate, maximized over the permissible ring-radius ratios, is shown for 16-APSK and 32-APSK in Fig. \ref{Fig_Apsk16_32Cap}.   

Next, we compute information rates when shaping is used.  For each constellation $\mathcal X$, we consider all shaping codes with $n_s \leq 20$ and $k_s \leq 10$.  Larger values are not considered due to their high decoding complexity (which is exponential in $k_s$).  For each shaping code, we determine the corresponding value of $p_0$.  For a given number of shaping bits $g$ and the shaping strategy given by Section \ref{Sec_Shaping}, we determine the pmf $p(x)$ and compute the corresponding information rate.  The total number of distinct $p_0$ that we consider is 121, and we repeat this exercise for each permissible value of $\gamma$. For each $M$ and $g$, we determine the maximum information rate at each value of $\mathcal E_s/N_0$, where the rate is maximized over both the permissible $p_0$ and $\gamma$.

\begin{figure}
\centering
\vspace{-0.07cm}
\includegraphics[width=3.45in,height=2.72in]{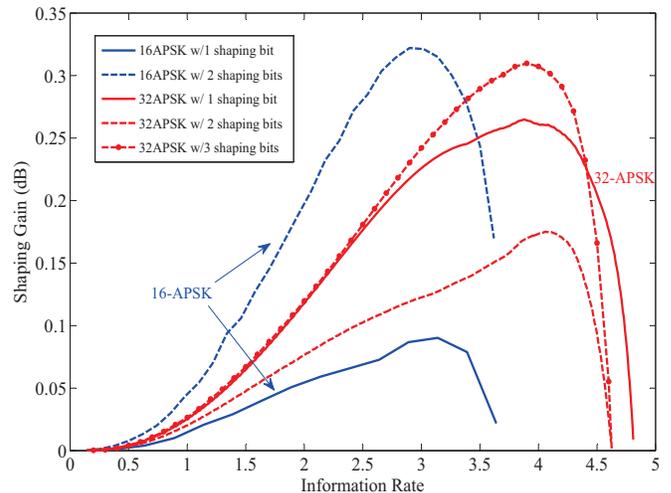}
\vspace{-0.72cm}
\caption{Shaping gain (in dB) of 16-APSK and 32-APSK in AWGN as a function of the overall information rate $R$ (in bits/symbol).  }
 \label{Fig_ShapingGain}
\vspace{-0.25cm}
\end{figure}

Based on our search, we have found that shaping gains of up to $0.32$ dB are achievable.  Fig. \ref{Fig_Apsk16_32Cap} shows the achievable information rates with shaping for 16- and 32-APSK in AWGN, optimized over the shaping code and ring-radius ratios. The shaping gain, which is the dB difference between the capacity curves of the uniform and shaped systems, is shown in Fig. \ref{Fig_ShapingGain}  for different values of $g$. The results for 16-APSK show the benefit of using two shaping bits over just one shaping bit.  From these curves, it is clear that if shaping is used at all, then it is advisable to use two shaping bits.  The results for 32-APSK show that while three shaping bits is better than just one, using two shaping bits is actually worse than one.  This is because using just one shaping bit segments the constellation into two natural subconstellations: One containing the two innermost rings and the other containing just the outer ring.  On the other hand, two shaping bits creates an awkward partitioning that results in symbols in the middle ring being picked with different probabilities.  Thus, it is not advisable to use two shaping bits in the 32-APSK case.  Furthermore, the incremental gain of using three shaping bits over one shaping bit is negligible (see the leftmost pair of curves in Fig. \ref{Fig_Apsk16_32Cap}), and it is recommended that 32-APSK systems use just one shaping bit.  This result is intuitive: with one shaping bit, the system will simply choose from the outer ring or the inner two rings.

The results from the optimization are tabulated in Table \ref{awgntable}.  For each value of $M$ and $g$, the table shows the rate $R$ for which shaping has the highest gain, along with the values of $p_0$ and $\gamma$ that achieve the gain.  The value of $\mathcal E_b/N_0$ for which the shaped constellation achieves rate $R$ is listed, where $\mathcal{E}_b/N_0 = (1/R)(\mathcal{E}_s/N_0)$ is the SNR per information bit.

\setlength{\tabcolsep}{4pt}
\renewcommand\arraystretch{1.25}
\begin{table}[t]
 \centering
    \caption{Minimum required ${\mathcal E}_b/N_0$ in AWGN for M-APSK with $g$ shaping bits.  The optimal $p_0$ and $\gamma$ are shown.}\label{awgntable}
  \vspace{-0.2cm}

  \begin{tabular}{|c|c||c|c|c|c|c|}
  \hline
     M & $g$ & $R$ & $\mathcal E_b/N_0$ & $gain$ & $p_0$ &  $\gamma$ \\
  \hline
   \multirow{2}{*}{16} &   1 & 3.09 & 4.714 \mbox{dB} & 0.091 \mbox{dB} & 0.623 & 2.70 \\
  \cline{2-7}  &  2 & 2.95 & 4.077 \mbox{dB} & 0.322 \mbox{dB} & 0.688 & 2.57  \\
  \cline{1-7}
   \multirow{3}{*}{32} &   1 & 3.88 & 5.915 \mbox{dB} & 0.265 \mbox{dB} & 0.716 & \{2.64,4.64\} \\
   \cline{2-7}   &  2 & 4.06 & 6.517 \mbox{dB} & 0.175 \mbox{dB} & 0.623 & \{2.53,4.30\} \\
   \cline{2-7}   &  3 & 3.89 & 5.898 \mbox{dB} & 0.310 \mbox{dB} & 0.656 & \{2.53,4.30\}  \\
  \hline
  \end{tabular}
  \vspace{-0.3cm}
\end{table}

\begin{figure}
\centering
\includegraphics[width=3.4in]{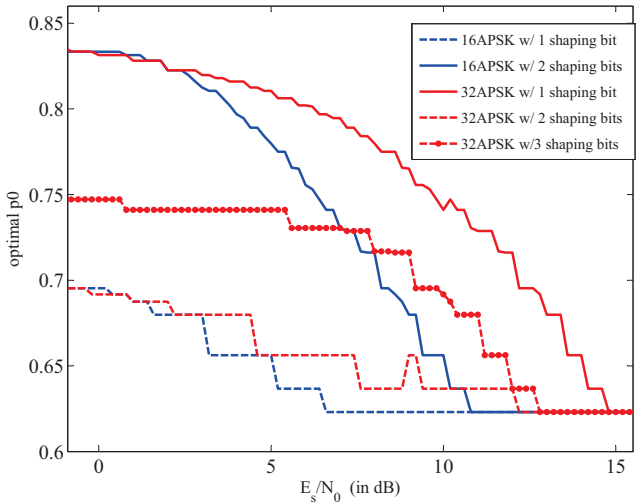}
\vspace{-0.4cm}
\caption{The value of $p_0$ that maximizes the information rate for nonuniform APSK in AWGN.  For 16-APSK, 1 and 2 shaping bits are used, while for 32-APSK, 1 through 3 shaping bits are used. } \label{Fig_OptimalP0}
\vspace{-0.35cm}
\end{figure}

Fig. \ref{Fig_OptimalP0} shows the value of $p_0$ that maximizes the information rate as a function of SNR.  The optimal $p_0$ decreases as the SNR increases, reaching a floor at $p_0 = 0.6230$.  This minimum value is a consequence of the constraints on $n_s$ and $k_s$ and corresponds to a $(n_s,k_s)= (11,10)$ shaping code.  While a $(k_s+1,k_s)$ code with larger $k_s$ could be used to obtain a lower $p_0$, the resulting shaping gain will be too small to merit the complexity of such a large code.

\subsection{PAPR Constraints}\label{Sec_PAPR}

The peak-to-average power ratio (PAPR) is a consideration when nonlinear power amplifiers must be used.  The PAPR is defined as
\begin{align}
  \mathsf{PAPR}
   = &
  \frac{  \underset{ x \in \mathcal X }{\max} \left\{ |x|^2 \right\} }
  { E \left[  |x|^2  \right] }
  =
    \frac{  \underset{ x \in \mathcal X }{\max} \left\{ |x|^2 \right\} }
  { \displaystyle \sum_{x \in \mathcal X} p(x) |x|^2  } \label{PAPR}
\end{align}

\begin{figure}
\centering
\includegraphics[width=3.4in]{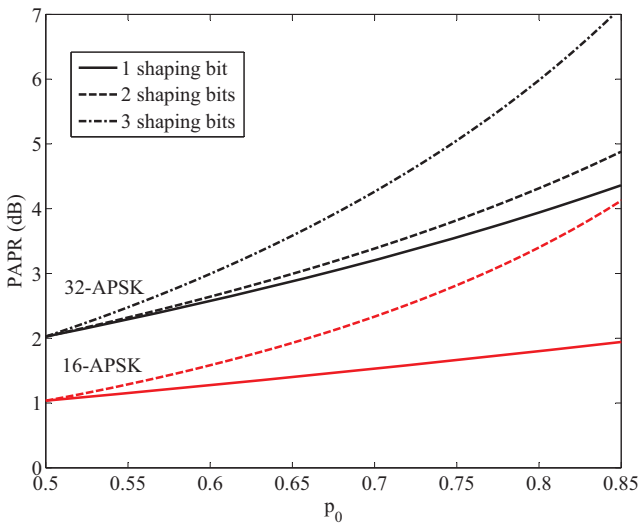}
\vspace{-0.35cm}
\caption{Peak-to-average power ratio (PAPR) of nonuniform 16-APSK and 32-APSK as a function of $p_0$ and the number of shaping bits. For 16-APSK, $\gamma = 2.57$, and for 32-APSK, $\gamma = \{2.64,4.64\}$.  } \label{Fig_PAPR}
\vspace{-0.55cm}
\end{figure}

\noindent For a fixed average energy, shaping spreads the signals further apart, which increases the numerator of (\ref{PAPR}). It follows that shaping will increase the PAPR for a particular constellation $\mathcal X$.  This behavior can be seen in Fig. \ref{Fig_PAPR}, which shows the relationship between PAPR and $p_0$ for the shaping strategies considered in this paper.

At first glance, it would seem that shaping will dramatically increase the PAPR.  This is true if we were to use the same $\mathcal X$ both with and without shaping.  However, when we performed our optimization, we let the ratio of ring radii vary among their permissible values.  When we did this, we found that the optimal ratio of ring radii were always smaller with shaping than without.  Furthermore, the PAPR is smaller for smaller ring-radii ratios.  Thus, in achieving the balance between the ratio of ring radii and the value of $p_0$, the PAPR increase is not significant.  For example, in the case of 16-APSK with 2 shaping bits, we found that the PAPR with shaping is 1.98 dB, while for the uniform case, which was optimized at the largest value of $\gamma$, was 1.11 dB, and thus only 0.86 dB additional PAPR is required when shaping.

%

\section{Results with LDPC Coding} \label{Sec_EXIT}

In this section, we consider the  AWGN bit-error performance of a system with the proposed shaping technique when an actual LDPC code is used.  As a running illustrative example, we consider a system that uses 32-APSK modulation operating at a coded rate of 3 bits/symbol. Initially, we pair the shaped system with a code from the DVB-S2 standard \cite{dvb:2009}.  However, the DVB-S2 codes were not optimized for use with the proposed shaping technique.  Thus, we next optimize the codes by using EXIT-based techniques similar to those proposed in \cite{brink:2004}, and show that the optimization provides an additional performance improvement. Finally, we make some observations on the complexity of the overall system, as compared with a uniform (non-shaped) system.

\subsection{Standardized DVB-S2 Codes}\label{dvbs2}
We begin by comparing the performance of shaped and uniform systems that use LDPC codes from the DVB-S2 standard \cite{dvb:2009}.  The APSK ring radii are selected to maximize the corresponding CM capacity, as described in Section \ref{Sec_Optimization}.  For both systems, the overall rate is $R=3$ bits/symbol.  To achieve this rate, the uniform system must use a rate $R_c=3/5$ LDPC code, and so the $(N_c,K_c) = (64\,800, 38 \, 880)$ code from the DVB-S2 standard is adopted.  The shaped system uses $g=1$ shaping bit per symbol, which was found in Section \ref{Sec_Optimization} to offer high shaping gain while avoiding the system complexity of using multiple shaping bits per symbol.  A rate $R_s =1/2$ shaping code is used with $(n_s,k_s) = (4,2)$, which provides a reasonably good $p_0$ while still allowing the standardized LDPC code rates $R_c$ from the DVB-S2 standard to be used. 
From (\ref{rate}), when $g=1$ and $R_s = 1/2$, the LDPC code rate must be $R_c = 2/3$ for the overall rate to remain $R=3$ bits/symbol, and so the shaped system uses the $(N_c,K_c) = (64\,800, 43 \, 200)$ code from the DVB-S2 standard.

The parameters used in the simulations are listed in Table  \ref{parametertable}.  The column marked $\gamma$ shows the ring-radii ratios used for both the uniform and shaped systems, while the LDPC and shaping code rates are listed in the columns marked $R_c$ and $R_s$, respectively (with $R_s=1$ indicating cases when no shaping is used).  For each system, we consider the performance of several LDPC codes.  For this subsection, only those LDPC codes in the rows marked ``standard'' in the ``LDPC code'' column are of interest.  
The column marked ``information-theoretic minimum'' ${\mathcal E}_b/N_0$ lists the CM-capacity bound; i.e., the minimum ${\mathcal E}_b/N_0$ for which the CM capacity is equal to $R$ (see Section \ref{Sec_Optimization} for a discussion about how the CM capacity is computed).

\setlength{\tabcolsep}{2pt}
\begin{table}[t]
  \centering
  \caption{Parameters used for the 32-APSK simulation.}\label{parametertable}
  \begin{tabular}{|c|c||c|c|c|c|c|}
  \hline
  & \multirow{3}{*}{$\gamma$} & \multirow{2}{*}{LDPC} &  \multirow{3}{*}{$R_c$} & \multirow{3}{*}{$R_s$}& information-  & $\mathcal E_b/N_0$ \\
  &  &  &   &  &   theoretic   & at BER  \\
    &  & code &   &  &   $\mathcal E_b/N_0$  & $10^{-5}$ \\
  \hline
       \multirow{2}{*}{Uniform} &    \multirow{2}{*}{\{2.64,4.64\}} & \small{standard} & 38880/64800  & 1    &   4.029 \mbox{dB}& 5.42 \mbox{dB}\\
  \cline{3-7}                &                                   &  \small{optimized} & 38880/64800  & 1  &  4.029 \mbox{dB}&  5.28 \mbox{dB}\\
  \cline{1-7}
           \multirow{3}{*}{Shaped} &   \multirow{3}{*}{\{2.64,4.64\}}  & \small{standard} & 43200/64800   & 2/4   &  3.829 \mbox{dB} & 4.96 \mbox{dB}\\
  \cline{3-7}                      &                                   & \small{optimized} & 43200/64800   & 2/4 &  3.829  \mbox{dB} & 4.80 \mbox{dB}\\
 \cline{3-7}                      &                                   & \small{optimized} & 41661/64806   & 2/3 &  3.789  \mbox{dB} & 4.62 \mbox{dB}\\
  \hline
  \end{tabular}
  \vspace{-0.25cm}
\end{table}

The uniform and shaped systems were simulated over an AWGN channel with 100 iterations of decoding.  For the uniform system, two receiver implementations are considered.  The first implementation is a {\em BICM} receiver, which passes the bit-likelihoods produced by the APSK demodulator into the LDPC decoder without any feedback from the decoder to the demodulator.   This is typical of most standard implementations of DVB-S2 receivers.  The second implementation is a {\em BICM-ID} receiver, which feeds back soft information from the LDPC decoder back to the APSK demodulator to allow it to refine its bit-likelihoods \cite{xie:vtc2009}.  The shaped system uses the iterative receiver described in Section \ref{Receiver}, which involves the iterative exchange of information among the APSK demodulator, shaping decoder, and LDPC decoder. 

\begin{figure}[t]
\centering
\includegraphics[width=3.4in]{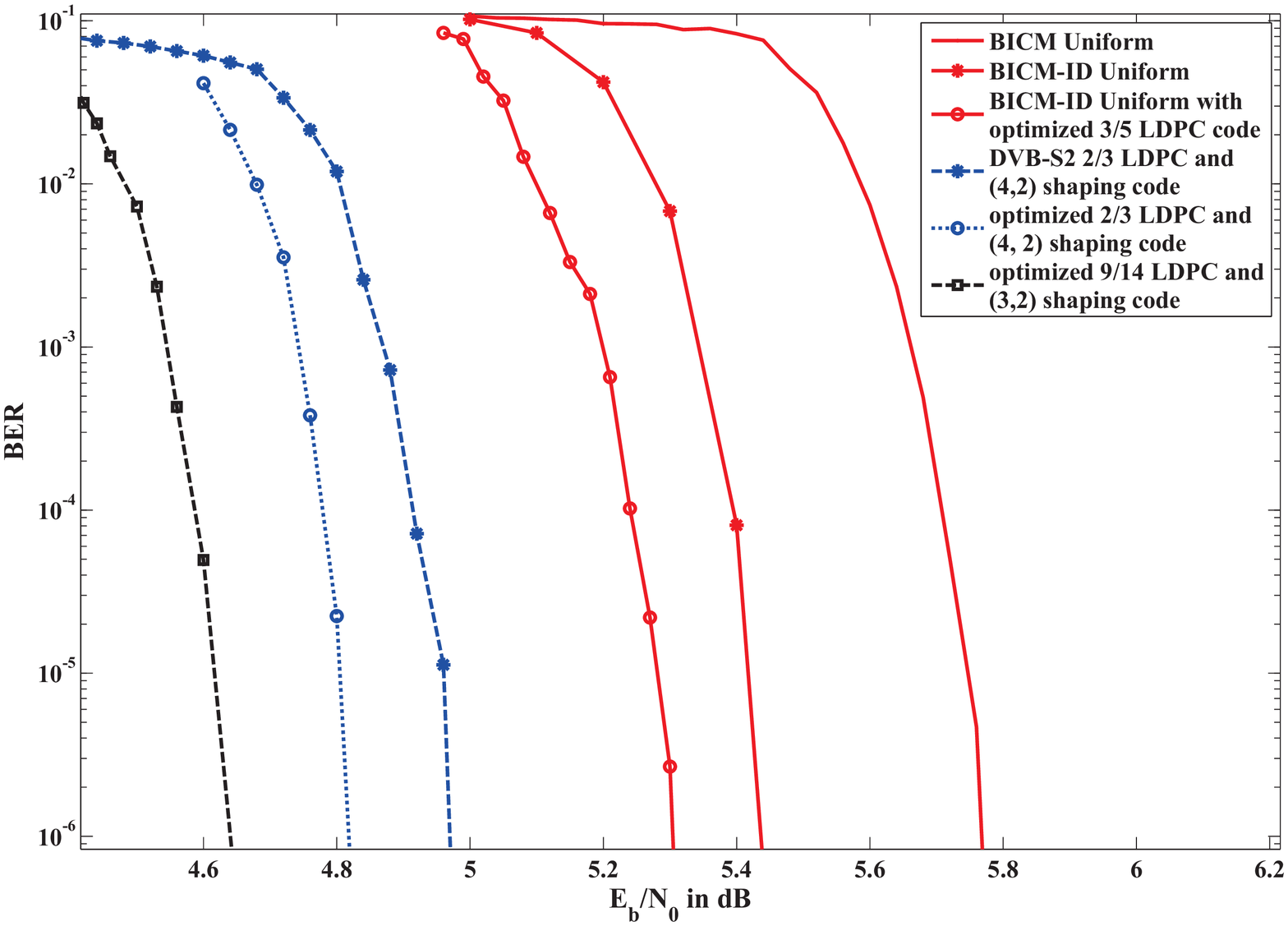}
\vspace{-0.2cm}
\caption{Bit-error rate of 32-APSK in AWGN at rate $R=3$ bits/symbol.  The solid lines are for the uniform system; from right to left, the curves show BICM system, BICM-ID system and BICM-ID system using optimized LDPC code. The dashed lines are for the shaped system with different combinations of LDPC codes and shaping codes. } \label{Fig_BER32AWGN}
\vspace{-0.5cm}
\end{figure}
Fig. \ref{Fig_BER32AWGN} shows the simulated bit-error rates of: (1) the uniform system with the rate $R_c = 3/5$ DVB-S2 standardized LDPC code and BICM reception (rightmost curve); (2) the same uniform system but with BICM-ID reception (second curve from right); and (3) the shaped system using the $R_c=2/3$ DVB-S2 standardized LDPC code and iterative receiver (third curve from left).  In addition, the last column of Table \ref{parametertable} shows the ${\mathcal E}_b/N_0$ required for the simulated BER to equal $10^{-5}$ with iterative decoding, which is 5.42 dB and 4.96 dB for the uniform and shaped system, respectively.  From Fig. \ref{Fig_BER32AWGN}, it is observed that the uniform system requires an ${\mathcal E}_b/N_0$ equal to 5.75 dB to achieve a BER equal to $10^{-5}$ when the BICM receiver is used.   Thus, the shaped system achieves a gain of 0.79 dB relative to the uniform system with BICM reception.  Of this gain, 0.33 dB can be attributed to iterative demodulation and decoding (i.e., using a a BICM-ID receiver) while the remaining 0.46 dB can be attributed to shaping and the use of the shaping code.  Notice that this gain is actually higher than the 0.2 dB shaping gain predicted from the information theory (see the difference between the corresponding entries in Table \ref{parametertable}).  Loosely speaking, this extra gain is due to an additional coding gain that arises when using the shaping code in an iterative receiver.  More precisely, and based on our EXIT-based analysis (discussed in the next subsection), we have observed that the inclusion of the shaping code in the system has the additional benefit of increasing the slope of the variable-node EXIT curve, thereby driving down the convergence threshold.

\subsection{Code Optimization for Uniform Modulation}\label{ldpcopt}
As described in \cite{Eroz:2006}, the LDPC codes used by DVB-S2 were designed for efficient memory access. They were not optimized with respect to the APSK modulation, and were certainly not designed with shaping in mind.  Both the APSK modulation and shaping may be taken into account by using the EXIT-based technique of \cite{brink:2004}, which will provide optimized degree distributions for the code.  In this and the next subsection, we describe how we have optimized the LDPC code's degree distributions and demonstrate the resulting BER improvement.

We begin by discussing how the EXIT-based technique of \cite{brink:2004} can be used to optimize the LDPC code used by the uniform APSK-modulated system.  Our goal here is to produce a redesigned LDPC code that has the same complexity as the code specified by DVB-S2, yet offers improved BER performance when used with APSK modulation.  Like the DVB-S2 code, we limit the LDPC code to be an extended irregular repeat accumulate (eIRA) code \cite{yang:2004}, which facilitates systematic encoding.  The parity-check matrix of such a code contains a dual-diagonal matrix as its last $n-k$ columns and an arbitrary matrix for its first $k$ columns.  We further limit the code to be {\em check regular}, and require every check node to have degree equal to the constant check-node degree $d_c$ used by the DVB-S2 standardized code of the same rate and length.  For instance, in the case of the $(N_c,K_c) = (64\,800, 38 \, 880)$ code, the check-node degree is a constant value of $d_c = 11$.

Since the check-node degree is fixed, the degree optimization problem is to select the optimal variable-node degrees.  We adopt the notation of \cite{brink:2004} and allow $D$ different variable-node degrees $d_{v,i}$, $i=1,...,D$.   The fraction of nodes having degree $d_{v,i}$ is denoted $a_i$, while the fraction of edges incident to variable nodes of degree $d_{v,i}$ is denoted $b_i$.  As in \cite{brink:2004} and the DVB-S2 standard, we limit the number of distinct variable-node degrees to $D=3$, starting with $d_{v,1}=2$.   We set a fraction $a_1=(n-k)/n$ of nodes to be of degree
$d_{v,1}=2$, corresponding to the dual-diagonal part of the $H$ matrix\footnote{In DVB-S2, there is a single degree-1 variable node corresponding to the last column of the $H$ matrix.  While our code design maintains this degree-1 node, we do not explicitly list the fraction of nodes connected to degree-1 nodes since it will approach zero for large $N_c$.}.  We then pick a pair $\{d_{v,2}, d_{v,3}\}$, where $d_{v,2}$ is selected to be either 3 or 4 and $d_{v,3}$ is an integer no greater than 25.

The optimization requires two EXIT curves to be generated, one that corresponds to the variable-node decoder (VND) and another that corresponds to the check-node decoder (CND).  The goal is then to fit the two curves together by picking appropriate variable-node degrees.  The VND curve characterizes not only the variable nodes of the LDPC code, but also the characteristics of the modulation.  The overall VND curve is created by first generating a transfer characteristic for just the modulation and its detector at the given  $\mathcal E_s/N_0$.  This is the {\em detector} characteristic $I_{E,\mathrm{DET} }( I_{A}, \mathcal E_s/N_0 )$, where the mutual information $I_{A}$ is computed between the demodulator's {\em a priori} inputs and the modulator's input bits.  When $I_A = 0$, $I_{E,\mathrm{DET} }$ is the BICM capacity of the modulation \cite{brink:2004,Caire:1998}. For APSK,  $I_{E,\mathrm{DET} }$ cannot be generated in closed form for nonzero $I_{A}$, and thus it is generated through Monte Carlo simulation under the assumption that the demodulator's {\em a priori} input is conditionally Gaussian.
The VND curve for degree-$d_v$ nodes is found from the detector characteristic using
\vspace{-0.25cm}
\begin{multline}\label{VND_COM}
I_{E, \mathrm{VND}} \left(  I_A, d_v, \mathcal E_s/N_0 \right) = \\
J \left( \sqrt{ (d_v-1) [ J^{-1}(I_{A})^2 ] + [J^{-1} (I_{E, \mathrm{DET}}( I_{A}, \mathcal E_s/N_0 ) )]^2 } \right)
\end{multline}
where the $J$-function is given in \cite{tenbrink:2001} and can be computed using the truncated series representation of \cite{torrieri:isrn2011}. Note that (\ref{VND_COM}) only gives the EXIT curve for a single variable-node degree $d_v$, and therefore represents the VND curve for a regular code.  In the case of an irregular LDPC code, the VND curve is found by using \cite{brink:2004}
\begin{equation}\label{VND_COM_irregular}
I_{E, \mathrm{VND}} \left(  I_A, \mathcal E_s/N_0 \right)  = \sum_{i=1}^{D} b_i \cdot
I_{E, \mathrm{VND}} \left(  I_A, d_{v,i}, \mathcal E_s/N_0 \right).
\end{equation}
The CND curve is found by using \cite{brink:2004}
\begin{equation}
I_{E, \mathrm{CND}}
\left(  I_A, d_c \right)
 =
1 - J\left(
\sqrt{d_c-1} \cdot J^{-1}(1-I_A)
\right)
\end{equation}
where $I_A$ is the mutual information at the input of the check nodes.  The EXIT chart is drawn by noting that the $I_{E, \mathrm{CND}}$ produced by the CND becomes the $I_{A}$ at the input to the VND (which we denote $I_{A, \mathrm{VND}}$), while the $I_{E, \mathrm{VND}}$ produced by the VND becomes the $I_{A}$ at the input to the CND (which we denote $I_{A, \mathrm{CND}}$).  The chart plots the VND and CND curves with $I_{A, \mathrm{VND}} = I_{E, \mathrm{CND}}$ on the horizontal axis and $I_{E, \mathrm{VND}} = I_{A, \mathrm{CND}}$ on the vertical axis.  For a given degree distribution, VND curves are generated for several $\mathcal E_s/N_0$ and the threshold is determined to be the value of $\mathcal E_s/N_0$ for which the VND and CND just barely touch.

By considering different combinations of $\{ d_{v,2}, d_{v,3} \}$ and the corresponding variable-node degree distributions that satisfy the check-node degree constraint, code designs were identified with low thresholds.  One code found to be better than the standard rate $R_c=3/5$ DVB-S2 code when used with 32-APSK has the following degree distribution:
\begin{center}
\begin{tabular}{ c c c }
  $d_{v,1} = 2$ & $a_1 = 0.40$ & $b_1 =  0.182$\\
  $d_{v,2} = 4$ & $a_2 = 0.52$ & $b_2 =  0.473$\\
  $d_{v,3} = 19$ & $a_3 = 0.08$ & $b_3 =  0.345$\\
\end{tabular}
\end{center}
The uniform 32-APSK system was simulated using a code with this degree distribution, and the resulting BER with BICM-ID reception is shown in Fig. \ref{Fig_BER32AWGN} (third curve from right; labeled ``BICM-ID uniform with optimized 3/5 LDPC code'').  This code is also listed as the second row of Table \ref{parametertable}, which indicates a gain of 0.14 dB relative to the DVB-S2 standard code at BER $10^{-5}$.

\subsection{Code Optimization for Shaped Modulation}\label{ldpcoptshape}
When the constellation is shaped, the shaping encoder (or decoder) is absorbed into the variable-node encoder (or decoder), and thus the contribution of the shaping code is taken into account by the VND curve.  The system effectively uses a larger constellation with high dimensionality that combines shaping with conventional modulation.  For instance, when $g=1$ and a $(N_s,K_s)$ shaping code is used, $K_s + (m-1) \times N_s$ bits are used to select a sequence of $N_s$ symbols.  Grouping these correlated symbols together, the resulting super-constellation contains $2^{K_s + (m-1)N_s}$ symbols, each represented by $2N_s$ real dimensions (or $N_s$ complex dimensions).  
Whereas generating the detector characteristic for the uniform case involves independently modulating symbols and measuring the mutual information between the modulator input and demodulator  output, the characteristic of the shaped modulation must be generated with the shaping taken into account.  In reference to Fig. \ref{Fig_SystemModel}, an unbiased independent bit sequence $\mathbf v$ is generated and passed through the pictured processing to produce the modulated sequence $\mathbf x$.  The received noisy symbols $\mathbf y$ are processed to produce the extrinsic information $L_e(\mathbf v)$, and the detector characteristic is found by computing the mutual information between $\mathbf v$ and $L_e(\mathbf v)$.

\begin{figure}
\centering
\includegraphics[width=3.4in]{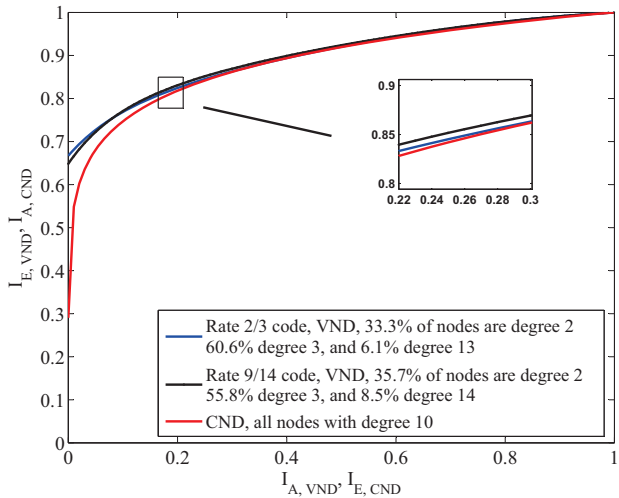}
\vspace{-0.3cm}
\caption{EXIT chart for BICM-ID using shaped 32-APSK and overall system rate R=3 in AWGN at $\mathcal E_b/N_0 = 4.73$ dB. } \label{Fig_exit}
\vspace{-0.5cm}
\end{figure}

Aside from accounting for the shaping code in the detector characteristic, the EXIT curves are generated exactly the same way as for the uniform case, and the variable-degree distributions are optimized using the same methodology.  For operation at 3 bits/symbol, a system that uses a $(N_s,K_s)=(4,2)$  and rate $R_c=2/3$ LDPC code was optimized, and the resulting degree distribution was found:
\begin{center}
\begin{tabular}{ c c c }
  $d_{v,1} = 2$ & $a_1 = 0.333$ & $b_1 =  0.200$\\
  $d_{v,2} = 3$ & $a_2 = 0.606$ & $b_2 =  0.546$\\
  $d_{v,3} = 14$ & $a_3 = 0.061$ & $b_3 =  0.254$\\
\end{tabular}
\end{center}
Like the rate-2/3 DVB-S2 code, this code has a constant check-node degree of $d_c = 10$.  The VND and CND curves for this LDPC code with  shaped 32-APSK modulation are shown in the EXIT chart of Fig. \ref{Fig_exit} at $\mathcal E_b/N_0 = 4.73$ dB.  The shaped system with was simulated using an LDPC code with this degree distribution, and the BER curve is shown in Fig. \ref{Fig_BER32AWGN} (second curve from left).  This code is also listed as the fourth row of Table \ref{parametertable}, which indicates a gain of $0.16$ dB compared with the system that uses the same shaping code along with the standardized rate $R_c=2/3$ DVB-S2 code.

In the previous examples, the code rate $R_c$ is chosen from the set of code rates found in the DVB-S2 standard.   However, due to (\ref{rate}), this limits the set of possible shaping-code rates for a given $R$, and may result in the use of a suboptimal $p_0$.  For instance, at an overall rate $R=3$ bits/symbol and $g=1$ shaping bit, then $R_c = 2/3$ and $R_s = 1/2$.  Using the $(n_s,k_s)=(4,2)$ shaping code will satisfy this requirement, but the resulting value of $p_0$ is $0.8125$, which differs from the optimal value of $p_0=0.716$ identified in Section \ref{Sec_Optimization}.  A $(3,2)$ shaping code gives a $p_0=0.75$, which is closer to the optimal, but requires LDPC code rate $R_c=9/14$ to achieve an overall rate of $R=3$ bits/symbol.  
While the standardized DVB-S2 codes do not support this code rate, a new code can be designed for this rate by using the EXIT-based techniques presented in this section, resulting in the degree distribution:
\begin{center}
\begin{tabular}{ c c c }
  $d_{v,1} = 2$ & $a_1 = 0.357$ & $b_1 =  0.200$\\
  $d_{v,2} = 3$ & $a_2 = 0.558$ & $b_2 =  0.469$\\
  $d_{v,3} = 14$ & $a_3 = 0.085$ & $b_3 =  0.331$\\
\end{tabular}
\end{center}
As with the rate-2/3 codes, the code has a constant check-node degree of $d_c=10$.  The VND curve for this code is also shown in Fig. \ref{Fig_exit}, and the BER performance is shown in Fig. \ref{Fig_BER32AWGN} (leftmost curve).  The code is listed as the last row of Table \ref{parametertable}, which indicates a gain of $0.18$ dB at BER $10^{-5}$ compared with the system that uses the optimized rate $R_c=2/3$ shaping code along with a $(4,2)$ shaping code.

\subsection{Complexity Considerations}\label{complexity}

\begin{figure}
\centering
\includegraphics[width=3.4in]{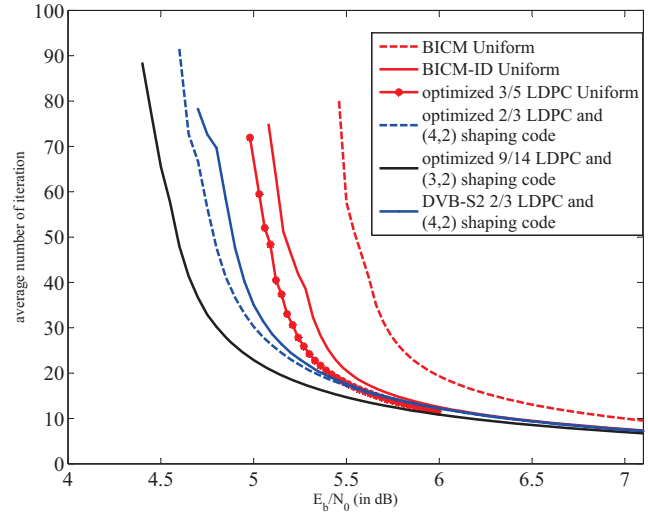}
\vspace{-0.25cm}
\caption{Average number of BICM-ID global iterations required for correcting all codeword errors.  } \label{Fig_avgIter}
\vspace{-0.5cm}
\end{figure}

Instead of using shaping as a way to improve energy efficiency, it can be used as a way to reduce the required receiver complexity.  This feature can be seen in Fig. \ref{Fig_avgIter}, which shows the average number of iterations required for the shaped and uniform systems to converge (i.e., correct all errors in a frame) in AWGN channel for the same cases whose BER curves were given in Fig. \ref{Fig_BER32AWGN}. This is a useful metric when the receiver operates using an automatic halting mechanism, for instance, if it performs syndrome-based error detection after each iteration. Notice that, at a given $\mathcal E_b/N_0$, the shaped system needs fewer iterations.  While the per-iteration complexity of the shaped system is higher than that of the uniform system, the need for fewer iterations may result in a lower {\em overall} complexity when early halting techniques are used. For instance, at $\mathcal E_b/N_0 = 5.4$ dB, which is a typical operating point for the unshaped system, an average of $25.3$ iterations are required in AWGN for the unshaped system with BICM-ID reception. At the same operating point, the shaped system with a DVB-S2 standard LDPC code and an optimized LDPC code only requires about $18$ and $15$ iterations to converge, respectively.

The overall and per-iteration complexity of the two systems can be compared by relating the actual execution time of their implementations.  The simulation results were produced with software implementations of the systems running on a PC computer with a single-core $3.40$ GHz Pentium $4$ processor and the Windows XP operating system.   For each system, the simulation's execution time, total number of data bits simulated, and total number of decoder iterations executed were logged.  From this information, the throughput of the implementation was quantified in units of bit$\cdot$iterations per second (bps$\cdot$iteration). For the uniform system with rate-3/5 DVB-S2 standardized coding, the processing rate is $2036.5$ bps$\cdot$iteration, while for the shaping system, the processing rate when using the DVB-S2 standard code is $1805.7$ bps$\cdot$iteration. The processing rate for optimized LDPC code with rate $2/3$ and $9/14$ is $1698.3$ and $1627.1$ bps$\cdot$iteration, respectively.  While shaping reduces the per-iteration throughput, the overall throughput may be increased due to the need for fewer iterations.   For instance, using the previously mentioned example operating point of $E_b/N_0 = 5.4$ dB over the AWGN channel, the uniform system has an average throughput of $78.1$ bps while the shaped system's throughput using the DVB-S2 standardized LDPC code is $100$ bps and for the optimized code is around $108$ bps.

\section{Conclusion} \label{Sec_Conclusion}

The combination of APSK modulation and LDPC coding is well suited for the constellation shaping technique promoted in this paper.  To design such a system, the probability $p_0$ of the shaping code and the ring radii of the APSK modulation should first be optimized using the CM capacity.   Next, the LDPC code should be optimized by using EXIT charts to select the variable-node degree distributions (assuming a constant check-node degree).  The system will require an iterative receiver, which iteratively demodulates the APSK signal, decodes the shaping code, and decodes the LDPC code.  When the $p_0$ is appropriately selected and the LDPC code is optimized, the resulting system with iterative reception outperforms a standard DVB-S2 system (which does not use iterative demodulation and decoding) by over 1 dB in AWGN with 32-APSK at a rate of 3 bits/symbol.  This gain is due to three factors: (1) using iterative reception, (2) shaping, and (3) optimization of the LDPC code.  The relative gain of each of these three factors is 0.33 dB, 0.46 dB, and 0.34 dB, respectively.   This is competitive with the shaping gains previously found for convolutionally coded QAM \cite{legoff:2006} and turbo-coded PAM \cite{legoff:2007}.  While shaping increases the required complexity per iteration, the need for fewer iterations opens up the opportunity for reduced overall complexity.  While we constrain the number of symbols per ring to match the numbers required DVB-S2, other configurations could be considered.  However, we note that the proposed system will work best with a single shaping bit if the number of symbols on the outer ring is equal to the number of symbols in the inner rings, as is the case for the 32-APSK configuration specified by DVB-S2.

\vspace{-0.25cm}
\section*{Acknowledgements} \label{Sec_Conclusion}
The authors would like to thank Dr. Don Torrieri for his editorial comments.
\vspace{-0.15cm}

\bibliographystyle{ieeetr}
\bibliography{proposal}

\vspace{-0.5cm}
\ifpdf
  \begin{IEEEbiography}{Matthew C. Valenti}
\else
  \begin{IEEEbiography}[{\includegraphics[width=1in,height=1.25in,clip,keepaspectratio]{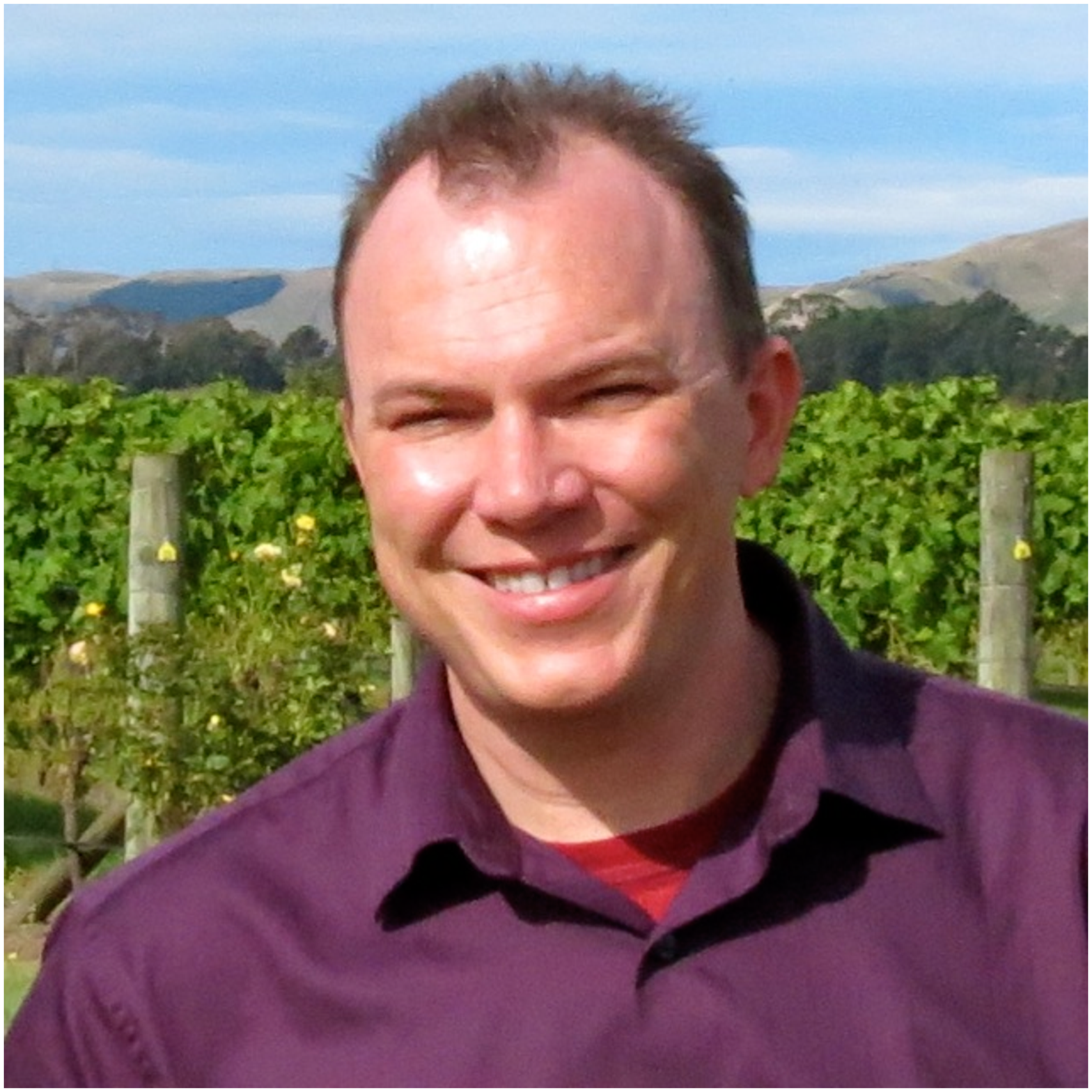}}]{Matthew C. Valenti}
\fi
is a Professor in the Lane Department of Computer Science and Electrical Engineering at West Virginia University. He holds BS and Ph.D. degrees in Electrical Engineering from Virginia Tech and a MS in Electrical Engineering from the Johns Hopkins University. From 1992 to 1995 he was an electronics engineer at the US Naval Research Laboratory.
He serves as an associate editor for {\em IEEE Wireless Communications Letters} and as Vice Chair of the Technical Program Committee for Globecom-2013.  Previously, he has served as a track or symposium co-chair for VTC-Fall-2007, ICC-2009, Milcom-2010, ICC-2011, and Milcom-2012, and has served as an editor for {\em IEEE Transactions on Wireless Communications} and {\em IEEE Transactions on Vehicular Technology}. His research interests are in the areas of communication theory, error correction coding, applied information theory, wireless networks, simulation, and secure high-performance computing.  His research is funded by the NSF and DoD.  He is registered as a Professional Engineer in the State of West Virginia.
\end{IEEEbiography}

\vspace{-0.5cm}
\ifpdf
  \begin{IEEEbiography}{Xingyu Xiang}
\else
  \begin{IEEEbiography}[{\includegraphics[width=1in,height=1.25in,clip,keepaspectratio]{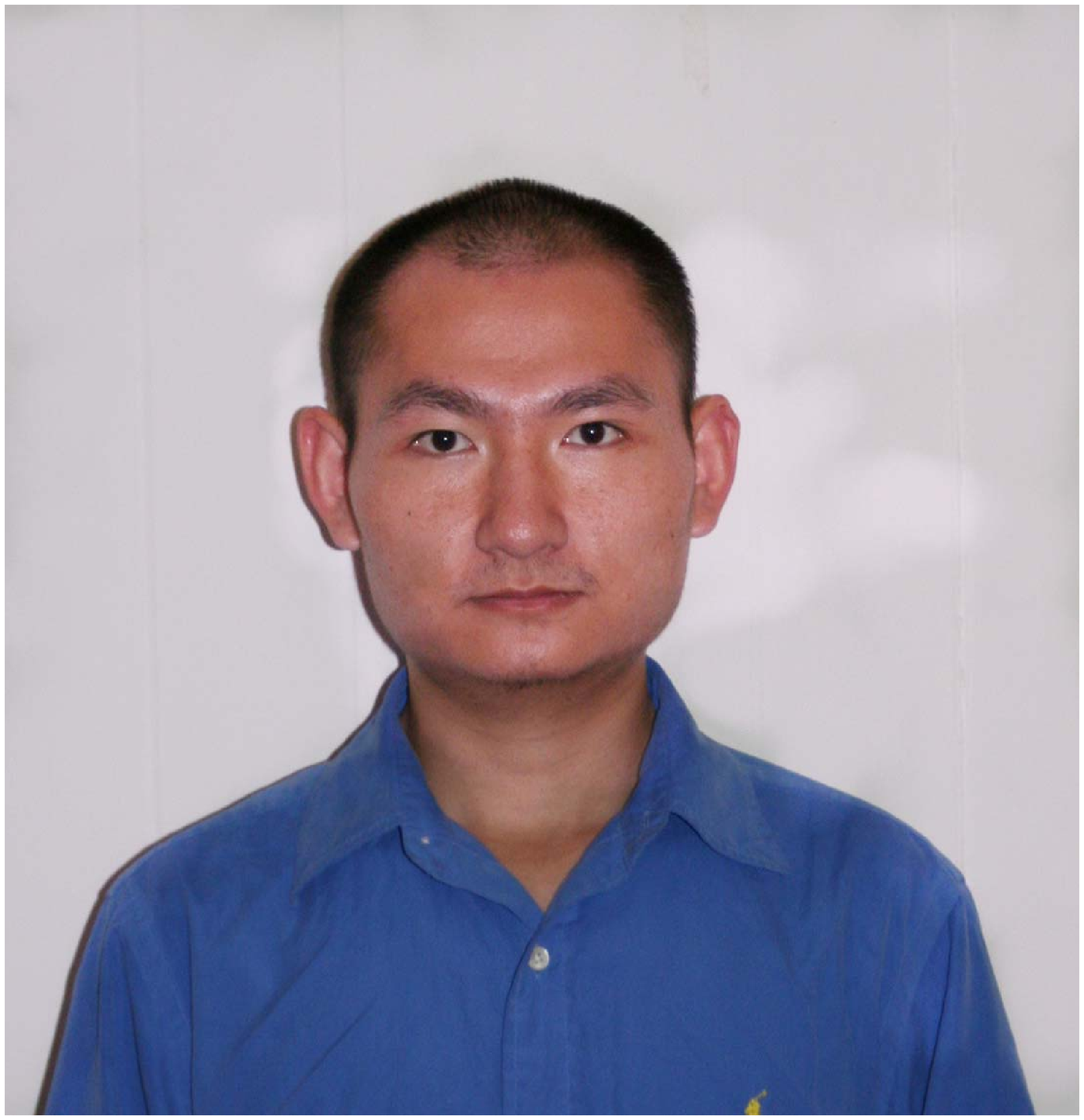}}]{Xingyu Xiang}
\fi
received the B.E. and M.S. degrees in electrical engineering from University of Electronic Science and Technology of China (UESTC), Chengdu, China in 2005 and 2008 respectively. He is currently a research assistant and a Ph.D. candidate in the Lane Department of Computer Science and Electrical Engineering at West Virginia University, Morgantown, WV. His research interests lie in the areas of information theory, channel coding, and communication signal processing.
\end{IEEEbiography}

\vfill

\end{document}